\documentclass[aps,pra,twocolumn,10pt]{revtex4-1}

\usepackage{graphicx}
\usepackage{amsmath,ulem}
\usepackage{color}
\usepackage{lineno}
\usepackage{units}
\usepackage{hyperref}
\usepackage{amssymb}
\usepackage[usenames,dvipsnames,svgnames,table]{xcolor}
\usepackage{siunitx}
\usepackage{wasysym}

\def\41K{$^{41}$K}
\def\39K{$^{39}$K}
\def\87Rb{$^{87}$Rb}

\def\ket#1{\left|#1\right\rangle}

\begin{document}
\title{Temperature dependence of an Efimov resonance in $^{39}\mathrm{K}$}
\author{L. J. Wacker$^{1, 2}$, N. B. J{\o}rgensen$^1$, K. T. Skalmstang$^1$, M. G. Skou$^1$, A. G. Volosniev$^3$ and J. J. Arlt$^1$}

\affiliation{$^1$ Institut for Fysik og Astronomi, Aarhus Universitet, 8000~Aarhus C, Denmark}
\affiliation{$^2$ Danish Fundamental Metrology, Kogle All\'{e} 5, 2970 H\o rsholm, DK}
\affiliation{$^3$ Institut f{\"u}r Kernphysik, Technische Universit{\"a}t Darmstadt, 64289 Darmstadt, Germany}
\date{\today}

\begin{abstract}
Ultracold atomic gases are an important testing ground for understanding few-body physics. In particular, these systems enable a detailed study of the Efimov effect. We use ultracold $^{39}\mathrm{K}$ to investigate the temperature dependence of an Efimov resonance. The shape and position of the observed resonance are analyzed by employing an empirical fit, and universal finite-temperature zero-range theory. Both procedures suggest that the resonance position shifts towards lower absolute scattering lengths when approaching the zero-temperature limit. We extrapolate this shift to obtain an estimate of the three-body parameter at zero temperature. A surprising finding of our study is that the resonance becomes less prominent at lower temperatures, which currently lacks a theoretical description and implies physical effects beyond available models. Finally, we present measurements performed near the Feshbach resonance center and discuss the prospects for observing the second Efimov resonance in $^{39}\mathrm{K}$.
\end{abstract}

\maketitle

\section{Introduction}
\label{sec:intro}

Three particles interacting via short-range two-body potentials possess an intricate universal spectrum of three-body bound states while the two-body subsystems are unbound~\cite{Efimov1970, Braaten2006, Braaten2007, Ferlaino2011, Ulmanis2016a, Naidon2017}. This feature is a cornerstone of few-body physics and is known as the Efimov effect.

Atomic vapors cooled to ultracold temperatures are an important tool for studying three-body systems. They provide unprecedented control and flexibility, e.g., Fesh\-bach resonances allow the two-body scattering length $a$ to be tuned to arbitrary values~\cite{Chin2010}. By choosing $a$ near the appearance of an Efimov state, three-body recombination losses are enhanced, and an Efimov resonance can be observed through loss spectroscopy~\cite{Fedichev1996,Burke1999,Braaten2006}. This loss signature of the Efimov physics allowed for the first unambiguous observation of an Efimov resonance~\cite{Kraemer2006}, and has since become the primary method of studying resonantly interacting three-body systems experimentally in both homo- and heteronuclear systems~\cite{Ottenstein2008, Huckans2009, Zaccanti2009, Pollack2009, Gross2009, Wenz2009, Gross2010, Berninger2011, Roy2013, Bloom2013, Huang2014a, Pires2014, Tung2014, Maier2015, Huang2015, Wacker2016, Ulmanis2016b, Ulmanis2016c, Johansen2017}.

A central property of Efimov states is their universal behavior across different atomic species and Feshbach resonances~\cite{Braaten2006, Braaten2007, Ferlaino2011, Berninger2011, Roy2013, Ulmanis2016a, Wacker2016, Naidon2017, Johansen2017}. The universal limit is reached in the ideal case of zero temperature and zero-range interactions. Here, the entire energy spectrum of three identical bosons is determined by the three-body parameter which fixes the location of the Efimov ground state, and by the universal scaling factor of approximately 22.7 which determines the spacing between the Efimov states. However, finite-temperature effects and finite-range interactions introduce modifications to this universal behavior. They drastically influence the appearance of Efimov resonances, hindering observations of consecutive resonances, and challenging the applicability of universality in few-body systems.

Several previous studies have considered the temperature dependence of Efimov resonances~\cite{Kraemer2006, Yamashita2007, Williams2009, Nakajima2011, Rem2013, Wang2014, Huang2014a, Huang2014b, Huang2015, Ulmanis2016b}, but the extent of systematic experimental investigations of single-component gases with focus on the temperature dependence is limited to a single study in a Cs ensemble~\cite{Huang2015}. By analyzing loss spectra obtained at different temperatures, it was found that the obtained Efimov resonance position has a temperature dependence, which cannot be accounted for by zero-range theory. 

Within this work, we study the temperature dependence of an Efimov resonances in a $^{39}\mathrm{K}$ sample. We introduce a preparation technique which ensures that the initial temperature of the ensemble is independent of the chosen interaction strength. The observed Efimov resonance changes its character with the temperature, and we analyze its appearance with two different methods. Both methods suggest that the apparent position of the resonance shifts towards smaller absolute values of the scattering length as the temperature is decreased. Contrary to theoretical expectations, the resonance becomes less pronounced at lower temperatures. By extrapolating the resonance position to zero temperature, we obtain an estimate of the three-body parameter for $^{39}\mathrm{K}$. Finally, we present measurements performed near the Feshbach resonance center and discuss the prospect for observing the second Efimov resonance in $^{39}\mathrm{K}$.

The rest of the paper is structured as follows. In Sec.~\ref{sec:theory}, we introduce the finite-temperature theory which will be used for characterizing the experimentally observed Efimov resonance. Section~\ref{sec:preparation} describes the experimental procedure for obtaining ultracold thermal samples. The method for evaluating the losses in these samples is presented in Sec.~\ref{sec:loss_eval}. The techniques for analyzing Efimov resonances are discussed in Sec.~\ref{sec:position_size} and the obtained results are provided in Sec.~\ref{sec:discussion}. Furthermore, in Sec.~\ref{sec:searchsecond}, we present measurements obtained at strong interactions and discuss the second Efimov resonance. Finally, we draw conclusions in Sec.~\ref{sec:conclusion}.

\section{Theory of Efimov resonances at finite temperatures}
\label{sec:theory}

The three-body loss of particles is described by the equation $\mathrm{d}n/\mathrm{d}t=-\alpha n^3$, where $n$ is the density of particles and $\alpha$ is the three-body recombination coefficient, which determines the probability for three particles to recombine~\cite{Braaten2006}. If $n(t)$ is known, this equation can be used to extract $\alpha$ from experimental data as described in Sec.~\ref{sec:loss_eval}. Here we briefly review how to relate $\alpha$ to the microscopic parameters. The analysis is based on the theory developed in Refs.~\cite{Platter2008, Rem2013}. However, instead of employing momentum space for calculations, we consistently use coordinate space. 

\begin{figure}[tb]
	\centering
	\includegraphics{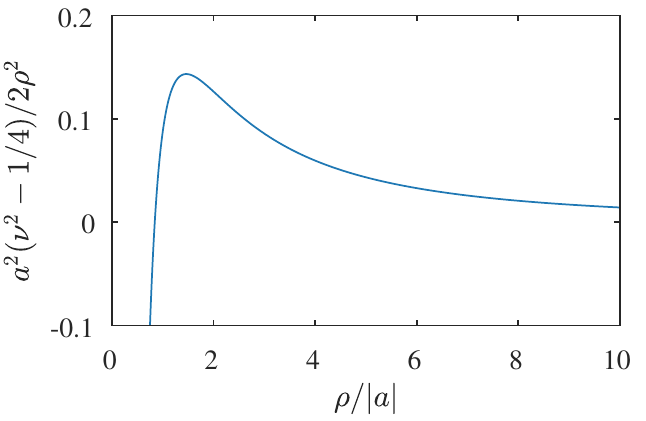}
\caption{Three-body hyperspherical potential, which supports Efimov states.}
	\label{fig:potential}
\end{figure}

Three spinless bosons are conveniently studied in the hyperspherical formalism~\cite{Nielsen2001}, where all relevant dynamics at low energies is described using a single differential equation
\begin{equation}
\left(-\frac{1}{2}\frac{\mathrm{d}^2}{\mathrm{d}\rho^2}+\frac{\nu^2(\rho)-1/4}{2\rho^2}-k^2\right)f(\rho)=0,
\label{eq:eq1}
\end{equation}
where $k^2=mE/\hbar^2$ with $E$ the energy of the system and $m$ the mass of a particle; $f(\rho)$ is the three-body wave function described in hyperspherical coordinates with $\rho=\sqrt{2/3}\sqrt{r_1^2+r_2^2+r_3^2-\mathbf{r_1}\cdot\mathbf{r_2}-\mathbf{r_2}\cdot\mathbf{r_3}-\mathbf{r_1}\cdot\mathbf{r_3}}$ (here $\mathbf{r_i}$ is the coordinate of the $i$th particle). The function $\nu(\rho)$ contains information about the two-body interactions. For a zero-range interaction potential~\footnote{A zero-range potential is the boundary condition for the two-body wave function at the origin: $\psi\sim 1/r-1/a$ as $r\to 0$.} the function $\nu(\rho)$ solves the transcendental equation
\begin{equation}
\frac{\nu\cos\left(\frac{\nu\pi}{2}\right)-\frac{8}{\sqrt{3}}\sin\left(\frac{\nu \pi}{6}\right)}{\sin\left(\frac{\nu \pi}{2}\right)}=-\sqrt{2}\frac{\rho}{|a|},
\label{eq:eq2}
\end{equation}  
where $a$ is the scattering length. Note that within this work we only consider $a<0$. 

The Schr{\"o}dinger equation~(\ref{eq:eq1}) reduces the complexity of the Efimov effect to the investigation of a simple one-body problem -- a particle in the $a^2(\nu^2-1/4)/(2\rho^2)$ potential. This potential is shown in Fig.~\ref{fig:potential}: it contains a barrier whose maximum $\simeq 0.14$ is located at $\rho/|a|\simeq 1.46$. It is repulsive for $\rho\to\infty$, whereas for $\rho\to 0$ it is attractive.

We now discuss the attractive region in more detail. For $\rho \to 0$ one of the solutions to~Eq.~(\ref{eq:eq2}) is imaginary $\nu_s=i s_0$, where $s_0\simeq 1.00624$. It leads to a (super) attractive $-1.2625/r^2$ potential in~Eq.~(\ref{eq:eq1}), which supports an infinite number of bound states with the ground state of infinite energy~\cite{Landau1965} -- the Thomas collapse~\cite{Thomas1935}. This collapse is due to the vanishing effective range parameter of the two-body potential; hence for small values of $a$ it is unphysical. For $|a|\to\infty$ the infinite tower of bound states is called the Efimov effect~\cite{Efimov1970}; the barrier is extended well beyond the two-body interaction range and Efimov states may be detected~\cite{Kraemer2006}.  

The existence of the Thomas collapse means that the Schr{\"o}dinger equation~(\ref{eq:eq1}) is ill defined: it has to be regularized at short distances. We do so by parametrizing the behavior of the three-body wave function at $\rho\to 0$~\cite{Efimov1979} for all $k$ as
\begin{equation}
f(\rho\to 0) \sim \sqrt{\rho}\left(A\rho^{\nu_s}+\rho^{-\nu_s}\right),
\label{eq:eq3}
\end{equation}
Here the parameter $A$ determines the short-range three-body physics at $k=0$. Note that the momentum $k$ plays a marginal role at $\rho\to 0$ (compared to the potential), and, therefore, we omit its effect here. All scattering observables can be now calculated from the wave function at large distances
\begin{equation}
f(\rho\to\infty)=H e^{-i \sqrt{2}k\rho}+G e^{i \sqrt{2}k\rho}.
\label{eq:eq4}
\end{equation}
To obtain the scattering amplitudes $H$ and $G$ numerically, one solves Eq.~(\ref{eq:eq1}) with the conditions~(\ref{eq:eq3}) at short distances and then fits the solution to the large-distance form given by Eq.~(\ref{eq:eq4}). To optimize this approach one can first solve the Schr{\"o}dinger equation with the following boundary conditions (cf.~\cite{Rem2013}),
\begin{align}
&f(\rho\to 0) \sim \sqrt{\rho}\left((k\rho)^{\nu_s}+s_{11}(k\rho)^{-\nu_s}\right), \\
&f(\rho\to\infty)=e^{i\sqrt{2}k\rho}, 
\end{align}
thus determining the function $s_{11}(k|a|)$. Note that the function $f$ here only has an outgoing flux at $\rho\to\infty$, whereas $f^*$ only has an incoming one. Once the function $s_{11}$ is known, the scattering amplitudes are easily computed for every value of the parameter $A$
\begin{equation}
\frac{G}{H}=\frac{s_{11}^*k^{\nu_s}-A k^{-\nu_s}}{s_{11} A k^{-\nu_s}-k^{\nu_s}}.
\end{equation}

To simulate the loss of particles we assume~\footnote{Another way to simulate a loss of particles is to introduce an imaginary part to the three-body potential~\cite{Sorensen2013}. We checked that this optical potential method also reproduces our data at high temperatures.} that $|A|<1$ (see~\cite{Braaten2006} and references therein), which means that some particles are lost close to $\rho=0$.  The recombination coefficient for a given momentum is then (cf.~\cite{Sorensen2013})
\begin{equation}
\alpha_k=36(2\pi)^2\sqrt{3}\frac{\hbar}{m k^4}\left(1-\left|\frac{G}{H}\right|^2\right),
\end{equation}
where $1-|G/H|^2$ determines the number of particles lost in the scattering governed by Eq.~(\ref{eq:eq1}). The prefactor connects this one-body problem to the three-body one. 
In terms of $A$, the parameter $\alpha_k$ is given by 
\begin{equation}
\alpha_k=36(2\pi)^2\sqrt{3}\frac{\hbar}{m k^4}\frac{(1-|A|^2)(1-|s_{11}|^2)}{|1-k^{-2\nu_s}s_{11} A|}.
\label{eq:eq9}
\end{equation}
To obtain the recombination coefficient for a fixed temperature we thermally average it assuming the Boltzmann distribution~\cite{DIncao2004}
\begin{equation}
\alpha=\frac{1}{2(k_\text{B} T)^3}\int \alpha_k E^2 e^{-E/k_\text{B}T}\mathrm{d}E,
\label{eq:eq10}
\end{equation}
where $k_\text{B}$ is Boltzmann's constant. 

\begin{figure}[tb]
	\centering
	\includegraphics{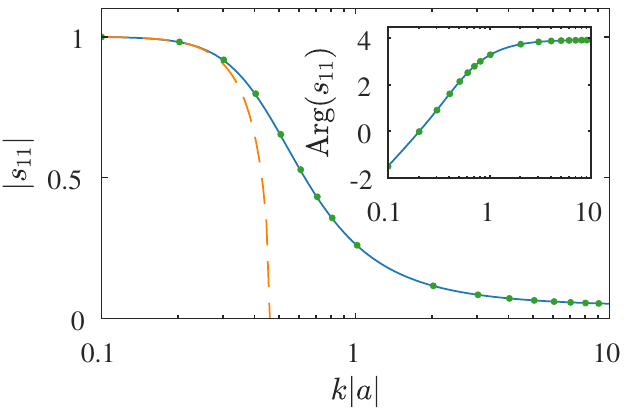}
\caption{Function $|s_{11}|(k|a|)$ calculated from Eq.~(\ref{eq:eq1}) shown as a solid blue line, and as green dots from Ref.~\cite{Rem2013}. The analytical limit for $k|a|\to 0$: $|s_{11}|^2\simeq 1-22.37 (ka)^4$, is shown as the dashed orange curve. The inset shows the argument of $s_{11}$, where the blue solid curve is calculated using Eq.~(\ref{eq:eq1}), and the green dots are from Ref.~\cite{Rem2013}.}
	\label{fig:s11}
\end{figure}

To calculate $\alpha$ we hence need to compute $s_{11}$ and specify $A$. The latter we write as $A=-R_0^{-2\nu_s}e^{-2\eta_-}$ (cf.~\cite{Rem2013}), where $\eta_-$ defines the recombination rate; note that the wave function vanishes at $\rho=R_0$ for $\eta_-=0$. The function $s_{11}(k|a|)$ has previously been calculated~\cite{Platter2008, Rem2013} using the Skornikov-Ter-Martirosyan equation~\footnote{Note that there is a one-to-one mapping from $s_{11}$ at $a<0$ onto $s_{11}$ at $a>0$~\cite{Rem2013}, and therefore the results of~\cite{Platter2008} obtained for $a>0$ also determine $s_{11}$ for $a<0$.}. Here we calculate it directly using the Schr{\"o}dinger equation~(\ref{eq:eq1}) -- we fix the boundary conditions at $\rho\to\infty$ and use a finite-difference method to calculate the function at $\rho\to 0$, which determines $s_{11}$. The function $s_{11}$ calculated in this way agrees well with previous results~\cite{Platter2008, Rem2013}. To give some insight into $s_{11}$, we plot it in Fig.~\ref{fig:s11}. Note that $|s_{11}|(0)=1$, i.e., transmission is not possible at zero energy. This limit follows directly from one-dimensional scattering theory. The behavior beyond this trivial limit can be obtained using the ideas of~\cite{Macek2005, Castin2011} as discussed in~\cite{Rem2013}. 

To relate $R_0$ to the standard three-body parameter $a_-$ we match $\alpha_k$ from Eq.~(\ref{eq:eq9}) to $\alpha_0$ derived in~\cite{Braaten2004, Braaten2006} 
\begin{equation}
\alpha_0=\frac{3\times 4590 \sinh(2\eta_-)}{\sin^2[s_0\ln(a/a_-)]+\sinh^2(\eta_-)}\frac{\hbar a^4}{m}.
\label{eq:eq11}
\end{equation}
We obtain $|a_-|= e^{(\delta-\pi n/2)/s_0} R_0$ and choose $n=1$, so $|a_{-}|\simeq 1.017 R_0$  (cf.~\cite{Rem2013, PetrovWerner2015}). Note that we used $s_{11}(x\to 0)=x^{2i s_0}e^{-2i\delta}$, where $\delta\simeq 1.588$~\cite{Macek2005, Rem2013}; moreover, we derived $|s_{11}(k|a|)|^2\simeq 1-22.37 (ka)^4$. 

The parameter $|a_-|$ is central to understanding the Efimov effect in ultracold atoms. It defines the scattering lengths at which $\alpha_0/a^4$ is maximal, at zero temperature. The universal properties of Efimov physics predict that $|a_{-}|$ lies within the interval $[8.27 R_\text{vdW},11.19 R_\text{vdW}]$~\cite{wang2012a}, where $R_\text{vdW}$ is the van der Waals length; see also~\cite{Zwerger2012, Sorensen2012,Naidon2014,Zwerger2018}. To date, there are no theories that relate $\eta_-$ to other microscopic parameters.

\begin{figure}[tb]
	\centering
	\includegraphics{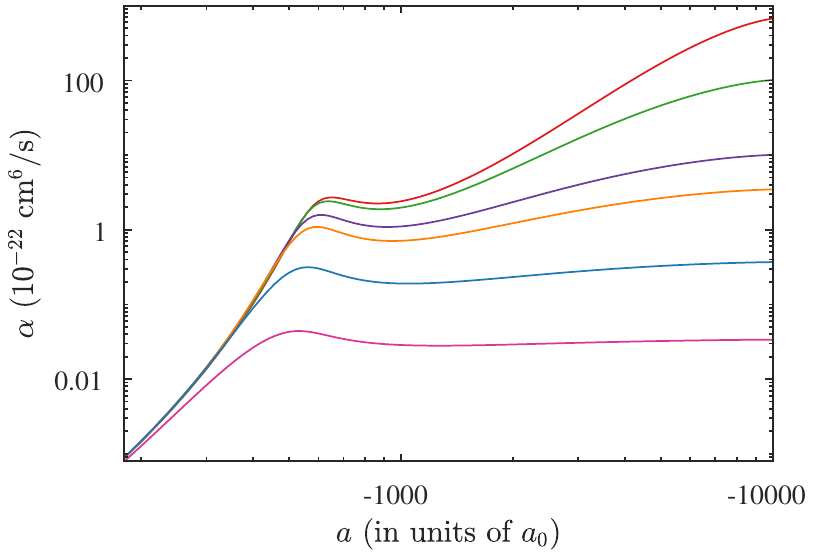}
\caption{Recombination coefficient $\alpha$ as a function of the scattering length $a$ for $T=5000,\; 1500,\; 500,\; 300,\;100$, and \SI{40}{nK} (bottom to top). Note that the apparent peak position shifts with temperature, even though all calculations were performed for the same value of $a_-$.}
	\label{fig:alpha}
\end{figure}

Figure~\ref{fig:alpha} shows $\alpha$ calculated for \39K at different temperatures. For the sake of argument, we have chosen $|a_{-}| = 600 a_0$ ($a_0$ is the Bohr radius) and $\eta_-=0.2$. At small values of $|a|$ the temperature effects are negligible and all curves coincide. At larger values of $|a|$ the finite-temperature effects become influential, they significantly alter $\alpha$ and the appearance of Efimov resonances, which are visible only for temperatures that allow a sizable portion of atoms to scatter at the energies below the height of the three-body barrier, i.e., $0.14 \hbar^2/(m |a|^2) \approx k_\text{B} \times \SI{1700}{nK}$. It is worth noting that, for increasing temperatures, the position of the recombination maximum shifts towards smaller $|a|$~\cite{DIncao2004, Kraemer2006, Jonsell2006, Yamashita2007}, which was also observed previously~\cite{Huang2015}. However, in~Ref.~\cite{Huang2015} this behavior had to be slightly corrected due to unknown finite-range effects, which led to the dependence of the three-body parameter $a_-$ on temperature. In future work, it will be interesting to incorporate finite-range corrections~\cite{Fedorov2003} into the theory to understand existing experimental data.

\section{Preparation and loss spectroscopy of ultracold \39K atoms}
\label{sec:preparation}

We study Efimov resonances experimentally by performing loss spectroscopy across a range of interaction strengths with \39K atoms prepared at different initial temperatures. 

The experiments were conducted using apparatus previously described in~\cite{Wacker2015}. Briefly summarized, a dual-species magneto-optical trap captures and cools $^{39}$K and $^{87}$Rb simultaneously in a glass cell. Subsequently, optical molasses and pumping is applied to both species, and they are captured in the $\ket{F=2, m_F=2}$ state by a magnetic quadropole trap. This trap mechanically transports the atoms to a different chamber, where they are loaded into another magnetic trap. Microwave radiation is applied to selectively evaporate $^{87}$Rb atoms, which cools $^{39}$K atoms sympathetically. All $^{87}$Rb atoms are evaporated, and the remaining $^{39}$K atoms are loaded into a crossed-beam optical dipole trap. Here, state preparation is carried out in two steps. Rapid adiabatic passages are performed to first transfer the atoms to the $\ket{2,-2}$ state, and finally to the $\ket{1,-1}$ state. The final evaporation is performed in the dipole trap by lowering the power of the two beams at a magnetic field of approximately \SI{41}{G}, where the rethermalization is enhanced due to the presence of the Feshbach resonance at \SI{33.64}{G}~\cite{Roy2013}. This resonance is also utilized later in the experimental procedure to investigate Efimov states.

\begin{figure}[tb]
	\centering
	\includegraphics{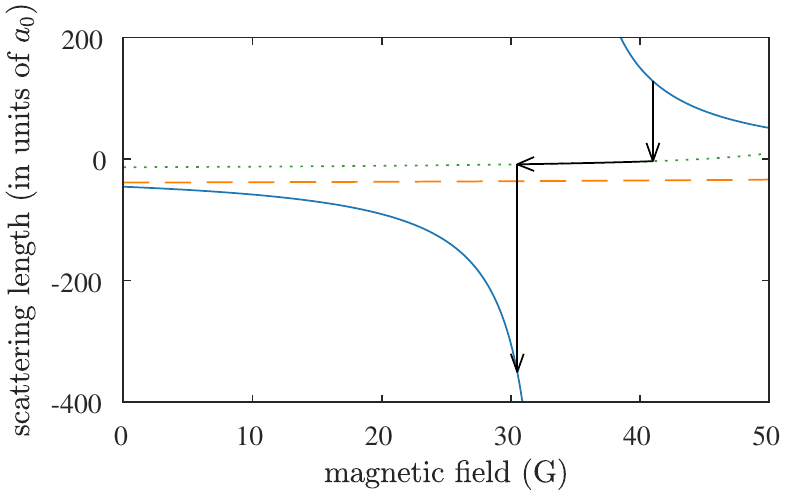}
	\caption{Scattering length (expressed in units of the Bohr radius $a_0$) as a function of the applied magnetic field for \39K atoms in the Zeeman states: $|F=1, m_F=-1\rangle$~(blue solid), and $|1,0\rangle$~(green dotted). The orange dashed curve shows the scattering length between the two states~\cite{Lysebo2010}. The arrow illustrates the preparation procedure to obtain ultracold ensembles at large interaction strengths for the observation of Efimov resonances.}
	\label{fig:Feshbach}
\end{figure}

An inherent experimental problem when accessing strong interactions is the finite speed at which a magnetic field can be changed. Before the target scattering length is reached, losses and dynamical processes can take place and introduce errors. To circumvent this inherent issue, we have developed a preparation procedure, which is shown schematically in Fig.~\ref{fig:Feshbach}.

When the evaporation in the $\ket{1,-1}$ state is complete and sufficiently low temperatures are reached, the atoms are transferred to the $\ket{1,0}$ state, which has a small negative scattering length. The magnetic field is then adjusted to a target value, and subsequently a wait time of \SI{0.5}{s} is added to ensure a stable field and complete rethermalization. Finally, the atoms are transferred back to the $\ket{1, -1}$~state, which initiates a loss measurement. This procedure avoids a direct exposure of the atoms to very large scattering lengths prior to the measurement and is essential to preserve the low temperatures achieved by evaporative cooling.

A loss measurement is performed by holding the sample for a variable time at a chosen interaction strength and releasing it from the trap afterwards. The magnetic field is turned off simultaneously with the release of the cloud. An absorption image is recorded after a total expansion time of \SI{20}{ms}, which allows the temperature and number of particles to be obtained.

\begin{figure}[tb]
	\centering
	\includegraphics{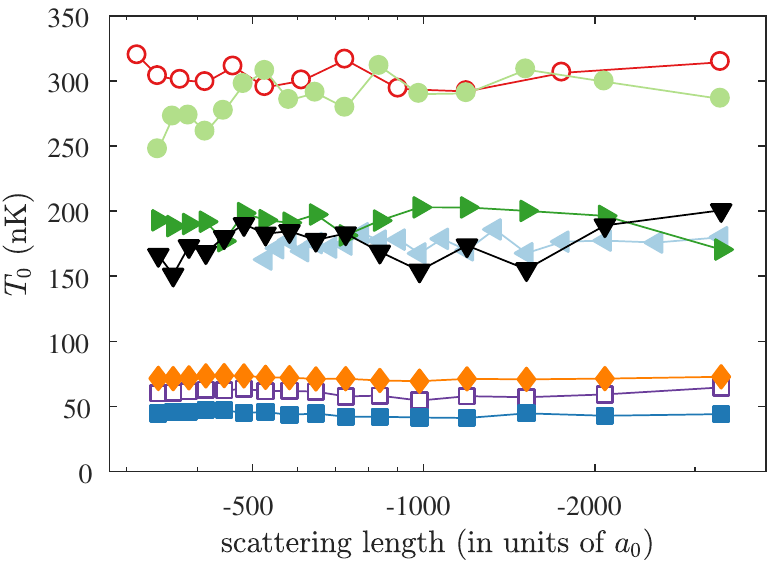}
\caption{Initial temperature of the atomic cloud prior to the loss measurements as a function of the scattering length. Each symbol refers to a specific data series, with more details provided in Table~\ref{tb:Tandn}. The initial temperature is obtained by performing a fit, as explained in Sec.~\ref{sec:loss_eval}.}
	\label{fig:T0_vs_a}
\end{figure}

To characterize the three-body loss, a series of decay measurements covering a range of interactions is performed. Multiple data series were acquired at various initial temperatures, which allows the temperature dependence of the Efimov resonance to be studied. The different initial temperatures are reached by evaporating and holding the atoms using various dipole trap configurations. In addition, the state preparation procedure into the interacting state was varied to test whether it had an influence on the observed Efimov resonance. The essential information on each data series is given in Table~\ref{tb:Tandn}.

In Fig.~\ref{fig:T0_vs_a} we show the initial temperature for all data series, obtained by performing a fit as described in the following section. The preparation of the ensemble through the noninteracting state clearly ensures a constant initial temperature across all interactions.

\begin{table*}[tb]
    \centering
        \caption{Sample parameters for all data sets. The symbols refer to Fig.~\ref{fig:T0_vs_a}. The parameters $\overline{T}_0$, $\overline{N}_0$, and $\overline{n}_0$ refer to the initial temperature, atom number, and density averaged across all decay measurements in a given data set. The state preparation procedure into the interacting state is also given: ``magnetic field ramp'' refers to the experiment in which the target scattering length is reached through a ramp of the magnetic field, instead of preparation from a weakly interacting state.} 
    \begin{tabular}{ l c c c c r }
        \hline \hline
        \rule{-4pt}{10pt} symbol & $\omega_x , \omega_y, \omega_z~(2\pi/\text{s})$ & $\overline{T}_{0}$ (nK) & $\overline{N}_{0}~(10^3)$ & $\overline{n}_{0}~(10^{11}/\text{cm}^{3})$ & state preparation procedure \\ \hline
        ${\color[rgb]{0.890,0.102,0.110}\fullmoon}$ & 83, 109, 83 & 304(9) & 34 & 7.7 & magnetic field ramp \\ 
        ${\color[rgb]{0.698,0.875,0.541}\newmoon}$ & 104, 142, 104 & 286(20) & 79 & 18.3 & $\pi$-pulse \\ 
        ${\color[rgb]{0.200,0.628,0.173}\blacktriangleright}$ & 86, 114, 86 & 192(9) & 20 & 10.0 & rapid adiabatic passage \\ 
        ${\color[rgb]{0.651,0.808,0.890}\blacktriangleleft}$ & 83, 109, 83 & 178(11) & 23 & 11.6 & rapid adiabatic passage \\ 
        ${\color[rgb]{0.000,0.000,0.000}\blacktriangledown}$ & 104, 142, 104 & 175(17) & 35 & 37.0 & $\pi$-pulse \\ 
        ${\color[rgb]{1.000,0.498,0.000}\blacklozenge}$ & 90.5, 28, 79 & 71.7(1.3) & 70 & 37.1 & $\pi$-pulse \\ 
        ${\color[rgb]{0.416,0.239,0.604}\square}$ & 89, 25, 75 & 60(3) & 49 & 27.8 & $\pi$-pulse \\ 
        ${\color[rgb]{0.122,0.471,0.706}\blacksquare}$ & 87, 21, 70 & 44(3) & 26 & 17.7 & $\pi$-pulse \\ 
        \hline \hline
    \end{tabular}
    \label{tb:Tandn}
\end{table*}

\section{Loss Evaluation}
\label{sec:loss_eval}
	
An Efimov state manifests itself experimentally as an increase of the three-body recombination coefficient $\alpha$ at a specific interaction strength. It is therefore necessary to carefully analyze atomic losses to characterize an Efimov resonance.

In a harmonic trap, three-body losses preferentially occur in the dense center of the ensemble. Since the average potential energy of atoms is lower here, three-body losses result in heating. At a specific interaction strength, both the change in the temperature $T$ and atom number $N$ thus have to be analyzed to obtain $\alpha$.

The time evolution of $T$ and $N$ can be described through the coupled differential equations $\text{d}N/\text{d}t = -\alpha \int n^3 ({\bf r}) \text{d} {\bf r}$ and  $\text{d}T/\text{d}t = \alpha T \int n^3 ({\bf r}) \text{d} {\bf r}/3N$. By assuming a Gaussian thermal distribution, the equations can be solved analytically to provide~\cite{Zaccanti2009,Roy2013}
	\begin{align}
	N(t)&= N_0\left(1+\frac{3\beta^2}{\sqrt{27}}\frac{N_0^2}{T_0^3} \alpha t\right)^{-1/3}\;,\\
	T(t)&= T_0 \left(1+\frac{3\beta^2}{\sqrt{27}}\frac{N_0^2}{T_0^3} \alpha t\right)^{1/9}\;,
	\end{align}
where $\beta=(m\bar{\omega}^2/2\pi k_\text{B})^{3/2}$, $m$ is the mass of \39K, $k_\text{B}$ is the Boltzmann constant, and $\bar{\omega} = (\omega_x \omega_y \omega_z)^{1/3}$ is the geometric mean of trapping frequencies. To obtain the three-body recombination coefficient from the decay measurements, these equations are simultaneously fitted to the atom number and temperature, which yields $\alpha$ as well as the initial atom number $N_0$ and temperature $T_0$. The temperatures shown in Fig.~\ref{fig:T0_vs_a} are obtained through this procedure.

\begin{figure}[tb]
	\centering
	\includegraphics{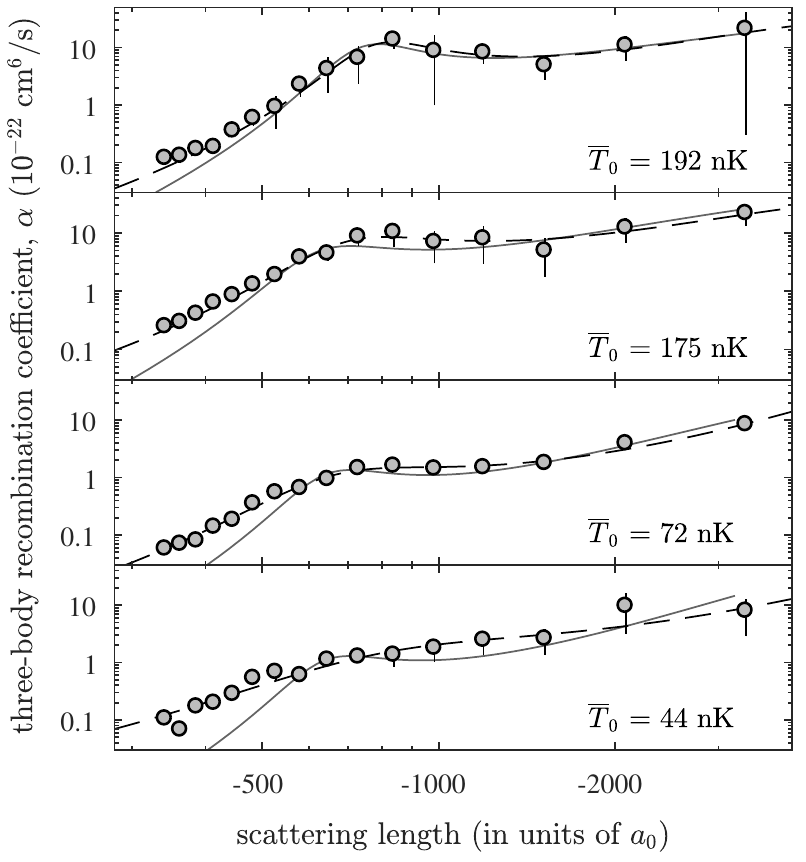}
	\caption{Three-body recombination coefficient as a function of the scattering length for four different temperatures. The average initial temperature $\overline{T}_{0}$ for each of the data sets is shown in the panels. Two different fits are performed to characterize the Efimov resonance: the dashed lines are empirical fits according to Eq.~(\ref{eq:Efimov_temp}) and the solid gray lines are fits using the finite-temperature theory [see Eq.~(\ref{eq:alpha_finite_temp_theory}) and the text].}
	\label{fig:three_body_coefficient}
\end{figure}

The three-body recombination coefficients for four different initial temperatures are shown in Fig.~\ref{fig:three_body_coefficient}. The magnetic field was converted into the scattering length using a previous characterization of the Feshbach resonance~\cite{Roy2013}. With increasing $|a|$, $\alpha$ tends to increase, as expected. Additionally, the ground-state Efimov resonance is present at approximately $-700 a_0$, which provides a local increase of $\alpha$. The position of the Efimov resonance is in close agreement with a previous observation~\cite{Roy2013}.

With decreasing temperature, the observed Efimov resonance changes. The apparent location of the resonance shifts towards a lower absolute value. Additionally, the resonance becomes less pronounced. At the lowest studied initial temperature of \SI{44}{nK}, the resonance is hardly distinguishable from the background slope.

The resonance behavior is in apparent disagreement with the zero-range theory presented in Sec.~\ref{sec:theory} and shown in Fig.~\ref{fig:alpha}. This points towards the presence of physics unaccounted for by the zero-range model, e.g., finite-range and many-body effects, and we will analyze the data from this perspective.

In Table~\ref{tb:Tandn}, information about the data sets is provided. The observed flattening of the resonance is not correlated with the density of the sample or the state preparation procedure, and we attribute this behavior to the change of the temperature.

\section{Efimov resonance characterization}
\label{sec:position_size}

In this section we present two approaches to quantitatively analyze the observed Efimov resonances. This allows for a detailed discussion of the shift and the unexpected suppression of the resonance at low temperatures.

\subsection{Analytic empirical fit}
\label{subsec:analytic_fit}

The three-body recombination coefficient $\alpha$  can be described analytically in the ideal limit of zero temperature and zero-range interactions~\cite{Braaten2006, Ferlaino2011}. However, in practice, finite temperature and finite-range interactions add upper and lower limits, alter the slope and change the Efimov resonance shape and position. Furthermore, systematic errors of the evaluated ensemble density can introduce inaccuracies. 

To obtain the apparent Efimov resonance location and width, we therefore perform an empirical fit
\begin{equation}
\alpha (a) = \frac{3 \times 4590\sinh(2 \eta_-)}{\sin^2[s_0 \ln(a/a_-)] + \sinh^2(\eta_-)} \frac{\hbar a_\text{e}^4}{m}  \left(\frac{a}{a_\text{e}}\right)^{n_\text{e}},
\label{eq:Efimov_periodic}
\end{equation}
which is similar to Eq.~(\ref{eq:eq11}). In addition, it contains the fitting parameters $n_\text{e}$ and $a_\text{e}$, which allow $\alpha$ to deviate from the predicted $a^4$ dependence and introduce an overall shift. Furthermore, we provide an upper constraint to $\alpha$, by introducing the effective three-body recombination coefficient
\begin{equation}
\alpha_\text{eff} (a) = \left( \frac{1}{\alpha (a)} + \frac{1}{\alpha_\text{max}} \right) ^{-1}
\label{eq:Efimov_temp}
\end{equation}
which is limited due to temperature according to 
\begin{equation}
\alpha_\text{max} = \frac{36 \sqrt{3} \pi^2 \hbar^5}{(k_\text{B} \overline{T}_0)^2 m^3},
\label{eq:Efimov_maximum}
\end{equation}
following previous work~\cite{Roy2013}. In Eq.~(\ref{eq:Efimov_temp}), the effective recombination rate is finite even at $|a|\to \infty$ when $T \neq 0$. Note that, unlike~\cite{Roy2013}, we do not fit $\alpha_\text{max}$. The fitting parameters are thus the apparent resonance position $a_-$, and the elasticity parameter of the trimer $\eta_-$, as well as the empirical parameters $n_\text{e}$ and $a_\text{e}$.

This fit is applied to the data sets, as shown in Fig.~\ref{fig:three_body_coefficient}. The fit describes the obtained data well across all temperatures, including the observed suppression of the resonance at low temperatures. For all of the data sets, we obtain that $n_\text{e} \approx 3$ and $a_\text{e}$ is of the order of $a_-$.

\subsection{Characterization through finite-temperature theory}
\label{subsec:theory_fit}

Based on the theory described in Sec.~\ref{sec:theory}, we also perform a numerical fit to obtain $a_-$ and $\eta_-$.

The fit is motivated by a clear separation of length scales in our experiment: the thermal length scale $\lambda_{\mathrm{th}}=h/\sqrt{2\pi m k_B T}$, the length scale associated with the trap $\sqrt{\hbar/m \bar \omega}$, and the interparticle distance $(1/n)^{1/3}$ are always considerably larger than $|a_-|$. Other relevant scales are much smaller than $|a_{-}|$: the van der Waals length is $R_\text{vdW}= 64.53 a_0$~\cite{Mitroy2015}, and the intrinsic length of the relevant Fesh\-bach resonance is $R^* = 23 a_0$~\cite{Roy2013}. Therefore, we use the parametrization in Eq.~(\ref{eq:eq10}) derived from the microscopic zero-range Hamiltonian to perform a fit. We write $\alpha$ as 
\begin{align}
\alpha & = \delta\frac{72\sqrt{3}\pi^2\hbar(1-e^{-4\eta_{-}})}{mt^3 a^2} \nonumber \\
	& \times \int_{0}^\infty \frac{(1-|s_{11}|^2)e^{\frac{-x^2}{ta^2}}}{\left|1+\left(\frac{x R_0}{|a|}\right)^{-2is_0}e^{-2i\eta_{-}}s_{11}\right|^2}x\mathrm{d}x,
\label{eq:alpha_finite_temp_theory}
\end{align}
where $t=\sqrt{m k_\text{B} T}/\hbar^2$. This expression is evaluated numerically with fitting parameters $\eta_-, R_0$ and $\delta$. The latter parameter accounts for the systematic errors of the experiment that originate from the evaluated ensemble density. Note that this fit contains less fitting parameters than the empirical fit. 

The fit is applied to the data as shown in Fig.~\ref{fig:three_body_coefficient}. Generally, the theoretical model fails to describe the observed three-body recombination for the four coldest samples. Under the conditions obtained at low temperatures, there are important physical effects that are not taken into account, such as finite-range effects. To minimize the influence of finite-range effects, we introduce a variable cut-off, which excludes data below a certain value of $|a|$. Varying the cut-off has minor influence on the obtained values of $a_-$ and $\eta_-$ for temperatures above $\SI{100}{nK}$. Below $\SI{100}{nK}$, the fit is cut-off dependent. The influence of the chosen cut-off value is used to estimate the uncertainty of the obtained fit parameters. For the results shown in Fig.~\ref{fig:three_body_coefficient} and treated later, a cut-off value of $500a_0$ was used.

\section{Evaluation of finite temperature behavior}
\label{sec:discussion}

The two different fitting procedures described above both provide measures of the apparent Efimov resonance position $a_-$ and the elasticity parameter $\eta_-$. In this section, we discuss these results to quantify finite-temperature effects.

Previous studies of Efimov resonances, were often based on fits that either use analytical expressions at zero temperature as in Eq.~(\ref{eq:Efimov_periodic}), or numerical calculations [cf.~Eq.~(\ref{eq:alpha_finite_temp_theory})] that take finite-temperature effects more formally into account. This section thus provides an inherent comparison of these different approaches to characterize Efimov resonances.

Note that the parameters $a_-$ and $\eta_-$ are defined in the zero-range finite-temperature theory, which does not describe our data well at the lowest temperatures. The obtained apparent values of $a_-$ and $\eta_-$ therefore represent a best approximation of the true values within each analysis technique. A true characterization would require a theoretical model which fully accounts for our observations.

\subsection{Efimov resonance position}

Generally, finite-temperature behavior arises when the thermal wavelength $\lambda_\text{th}$ becomes non-negligible in comparison to the three-body parameter $a_-$. In the zero-temperature case, $\lambda_\text{th}$ is infinite and does not influence the observed Efimov resonance; hence the observed resonance position $a_-$ is equivalent to the standard three-body parameter. However, as the temperature is increased the observed Efimov resonance is modified. Disregarding finite-range effects, the finite-temperature fit takes these modifications into account and should obtain the same $a_-$ independent of temperature. Any observed changes in $a_-$ from this evaluation therefore originate from aspects not taken into account, such as temperature-dependent finite-range effects. 

The empirical fit does not inherently take temperature effects into account. Any observed temperature dependence therefore also includes the temperature-dependent trends shown in Fig.~\ref{fig:alpha}.

\begin{figure}[tb]
	\centering
	\includegraphics{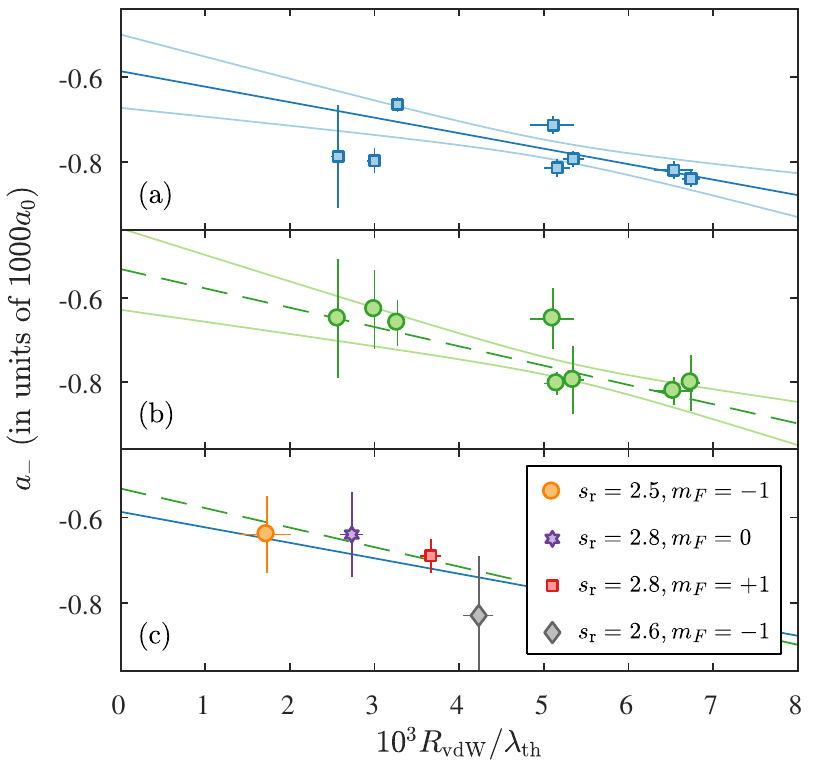}
	\caption{Apparent Efimov resonance location $a_-$ at various temperatures, parametrized by the van der Waals radius $R_\text{vdW}$ in units of the thermal wavelength $\lambda_\text{th}$. The blue squares in (a) are obtained using the empirical fit [see Eq.~(\ref{eq:Efimov_temp})], whereas the green circles in (b) are obtained using the finite-temperature theory fit [see Eq.~(\ref{eq:alpha_finite_temp_theory})]. The two fitting procedures are used to estimate the zero-temperature three-body parameter, shown as solid and dashed lines in (a) and (b), respectively. The corresponding standard errors are shown as faint lines. In (c), we compare the obtained fits to previous studies of Efimov resonances in \39K that used Feshbach resonances of similar strength $s_\text{r}$~\cite{Roy2013}. The legend provides the Feshbach resonance strength as well as the hyperfine state in which these measurements were performed.}
	\label{fig:position_vs_T0}
\end{figure}

The apparent values of $a_-$ obtained through the two fitting procedures are shown in Figs.~\ref{fig:position_vs_T0}(a) and \ref{fig:position_vs_T0}(b). The temperature is converted into the dimensionless parameter $R_\text{vdW}/\lambda_\text{th}$, which compares the relevant thermal wave length to $R_\text{vdW}$, where $\lambda_\text{th}$ is calculated using the average initial temperatures $\overline{T}_0$. For both evaluation methods, the value of $|a_-|$ shows a decreasing trend when the temperature is lowered. Note that this behavior is opposite to the apparent loss peak position shown in Fig.~\ref{fig:alpha}, predicted without finite-range effects. 

We estimate the zero-temperature value of the three-body parameter by performing a linear extrapolation to $\lambda_\text{th} \rightarrow \infty$~\cite{Huang2015}. In this linear fit, the measured errors of $a_-$ are included to weight the data points. The error bars in Fig.~\ref{fig:position_vs_T0}(b) are calculated by varying the cut-off value of $a$ by $100a_0$ in the finite-temperature fit. The change in obtained $a_-$ is added in quadrature to the standard fitting uncertainty. For the data obtained through the empirical fit, this procedure provides a zero-temperature value of $-587(86)a_0$, whereas the finite-temperature fit yields $-532(97)a_0$. The two results are within the errors of each other. We estimate a systematic error of the order of $50a_0$ due to an imprecision of the scattering length determination. The slopes obtained from the fits are similar, with $36\times10^3a_0$ for the empirical fit, and $46\times10^3a_0$ for the finite-temperature fit. In {Fig.~\ref{fig:position_vs_T0}}, we additionally show confidence intervals obtained by the linear fit, which display the uncertainty of our results. Even within these uncertainties, the slope of the linear fit is non-zero, indicating a temperature-dependent Efimov resonance position.

Universality predicts a three-body parameter within the interval $[8.27 R_\text{vdW},11.19 R_\text{vdW}]$ corresponding to $|a_{-}| \in [534 a_0,722 a_0]$. This is in agreement with both our experimental zero-temperature estimates, within the uncertainties. 

In Fig.~\ref{fig:position_vs_T0}(c), we compare the obtained results to the previous characterizations of Efimov resonances in \39K~\cite{Roy2013}. Since the Feshbach resonance strength $s_\text{r}$ influences the location of the observed Efimov resonance~\cite{Roy2013, Johansen2017}, we only show observations at Feshbach resonances of similar strengths. The Feshbach resonance used within this study has a strength of $s_\text{r} = 2.6$, whereas the resonances used for the data shown in Fig.~\ref{fig:position_vs_T0}(c) have strengths in the range $2.5$--$2.8$. These resonances also have similar values of $R^*$ in the range $22a_0$--$24 a_0$. These past observations compare well with both linear trends obtained from our two fitting methods. 

We now compare our observations to the previous systematic study of temperature effects in Cs~\cite{Huang2015}. In Ref.~\cite{Huang2015} a similar linear trend was observed for $|a_-|$, which decreased when the temperature was lowered. However, the slope is significantly stronger in our observations with \39K, indicating stronger finite-range effects. The observation in Cs is in closer agreement with the universal predictions of a three-body parameter, than our observations in \39K. It is possible that this is due to finite-range physics or the nature of the employed Feshbach resonance.

\subsection{Elasticity parameter and suppression of the Efimov resonance}

\begin{figure}[tb]
	\centering
	\includegraphics{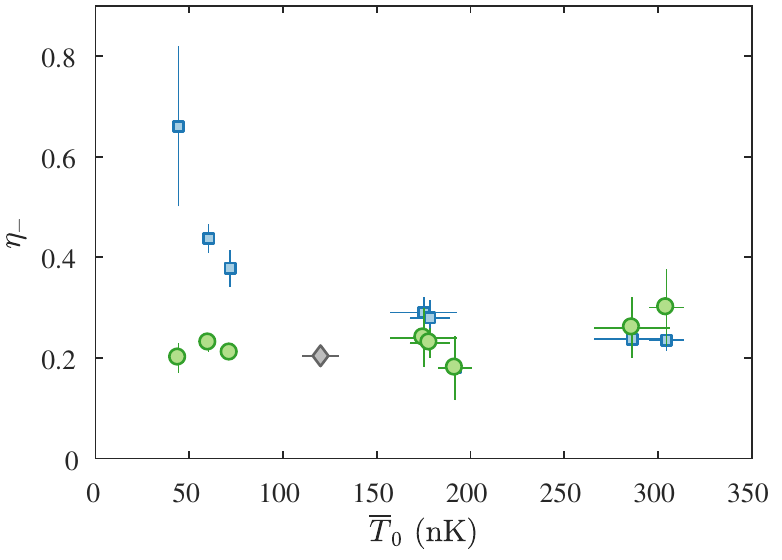}
	\caption{Elasticity parameter $\eta_{-}$ of the observed Efimov resonance at different initial average temperatures $\overline{T}_{0}$. The blue squares and green circles are obtained through empirical and finite-temperature fits, respectively. The gray diamond is the previous measurement from~\cite{Roy2013}.}
	\label{fig:decay_vs_T0}
\end{figure}

In Fig.~\ref{fig:decay_vs_T0}, we show the elasticity parameters $\eta_-$ at various initial sample temperatures, obtained from the two fitting procedures. The values of $\eta_-$ obtained from the empirical fit show an unexpected growth at low temperatures, which reflects the suppression of the resonance. The results obtained from the finite-temperature fit do not show the increase of $\eta_{-}$ at low temperatures. However, the finite-temperature fit generally agrees less well with the experimental data at low temperatures. These observations indicate that the suppression effect cannot be accounted for by the physics included in the finite-temperature theory and demonstrate the necessity of further theoretical investigations. 

There is no few-body mechanism, which is sensitive to the small temperature variation in the limit when the temperature is much smaller than any other energy scale of the problem. It is therefore relevant to consider many-body mechanisms to explain the suppression. In the experimental realization, the interparticle spacing is of the order of the thermal wavelength $\lambda_\text{th}$, and quantum statistics is therefore important. For the colder experimental samples, the critical temperature for Bose-Einstein condensation is in fact slightly above the actual initial temperatures.  Due to the experimental preparation through a weakly-interacting state with negative scattering length, it is possible that small Bose-Einstein condensates or solitons exist in the sample when the loss measurement is initiated. These many-body states may survive the hold time of \SI{0.5}{s} and we speculate that these many-body processes could influence the loss dynamics.

Another source of error which could potentially influence the loss dynamics is the presence of a few atoms not transferred into the target hyperfine state during the state-preparation procedure. However, it is not clear how the presence of a few weakly-interacting impurity atoms can significantly alter the rapid loss dynamics near the Efimov resonance.

\section{Second Efimov resonance}
\label{sec:searchsecond}

The results presented above provide a foundation for discussing the prospect of studying the resonance of the first excited Efimov state in \39K. For a single component quantum gas, this resonance has only been observed in Cs~\cite{Huang2014a}. This observation was performed at a temperature of approximately \SI{9}{nK}, which corresponds to $|a_-^{(1)}|/\lambda_\text{th} \approx 0.67$, where $a_-^{(1)}$ is the location of the excited state resonance. By considering \39K and assuming $- a_-^{(1)} = 22.7 \times 9.73 R_\text{vdW} = 14.2 \times 10^4 a_0$, a similar value of $|a_-^{(1)}|/\lambda_\text{th}$ is obtained at a temperature of approximately \SI{60}{nK}. Naively, it should therefore be possible to observe the excited state resonance in \39K at temperatures up to approximately \SI{60}{nK}.

\begin{figure}[tb]
	\centering
	\includegraphics{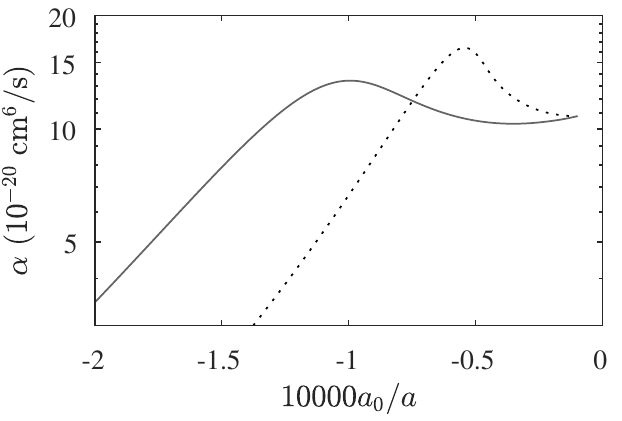}
	\caption{Theoretical comparison of the second Efimov resonance in Cs and \39K. The three-body recombination coefficient $\alpha$ of Cs is shown as the dotted black line, and as the gray line for \39K. The curves were calculated for sample temperatures($\eta_-, |a_-|$) of \SI{6.2}{nK}(0.10, 963$a_0$) for Cs and \SI{30}{nK}(0.25, 510$a_0$) for \39K. The Cs curve has been scaled by a factor of approximately 3.1 for an easier comparison.}
	\label{fig:K_Cs_compare}
\end{figure}

Based on our measurement of the Efimov ground state resonance, we model the first excited state Efimov resonance. The finite-temperature theory applied to Cs and \39K is shown in Fig.~\ref{fig:K_Cs_compare}, to compare the visibility of the previously observed Efimov resonance in Cs with a potential resonance in \39K. The theory predicts that under similar conditions the Cs resonance is the most distinct of the two. The \39K Efimov resonance is nevertheless distinguishable from a flat curve.

In an attempt to observe the excited state Efimov resonance, we performed a series of decay measurements near the Feshbach resonance center. The experiments were performed according to the description in Sec.~\ref{sec:preparation}, and the three-body recombination coefficient was obtained by fitting decay curves as described in Sec.~\ref{sec:loss_eval}. The initial average sample temperature across the range of magnetic fields was approximately \SI{20}{nK} rising to roughly \SI{42}{nK} during the measurement.

\begin{figure}[tb]
	\centering
	\includegraphics{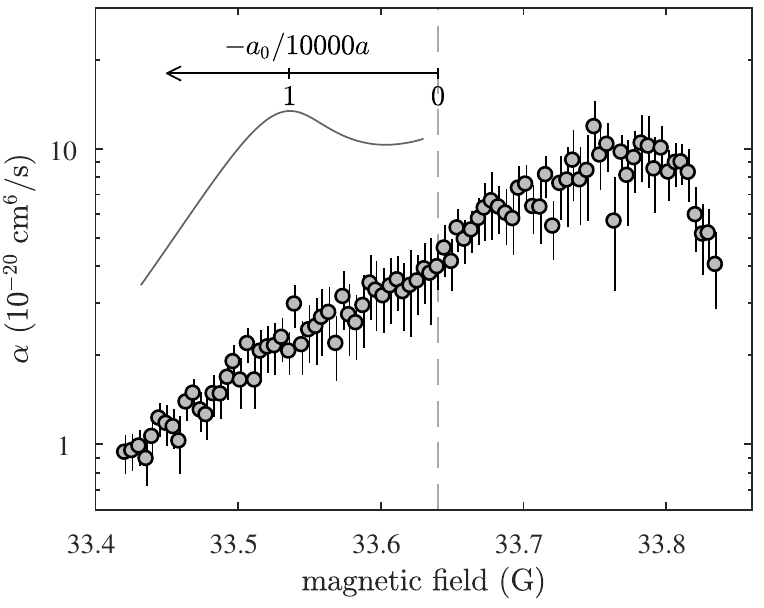}
	\caption{Three-body recombination coefficient $\alpha$ as a function of the magnetic field near the Feshbach resonance center. A theoretically predicated curve displaying the shape of the second Efimov resonance is shown (same as in Fig.~\ref{fig:K_Cs_compare}), assuming the Feshbach resonance center to be at the vertical dashed gray line. This assumption also provides the scattering length axis, shown near to the theoretical prediction. }
	\label{fig:Bfield_scan}
\end{figure}

The obtained three-body recombination coefficients $\alpha$ are shown in Fig.~\ref{fig:Bfield_scan}. Since an accurate calibration of the scattering length is not available near the Feshbach resonance center, we show $\alpha$ versus magnetic field. A loss maximum is observed at magnetic fields larger than the previously reported Feshbach resonance center. At strong positive scattering lengths, the presence of a weakly bound dimer state alters the loss dynamics and can indeed lead to a loss maximum not located at the resonance center, depending on experimental conditions~\cite{Zhang2011, Ulmanis2015}. In fact, the Feshbach resonance center was previously measured to be \SI{33.64(20)}{G} by locating the loss maximum~\cite{Roy2013}, but our data illustrates the deficiency of this method for accurate Fesh\-bach resonance characterization.

We now analyze the data in the context of observing an excited state Efimov resonance. In Fig.~\ref{fig:Bfield_scan} we show the theoretical prediction of the first excited state Efimov resonance, assuming the Feshbach resonance center to be at \SI{33.64}{G}. This allows us to calculate the scattering length axis, which is also shown. A different assumption about the Feshbach resonance center will shift the theoretical curve horizontally on the magnetic field axis. Based on a visual comparison between the data and the theoretical prediction, we do not observe any signatures of the second Efimov resonance.

There are several possible explanations as to why we do not observe the second Efimov resonance. Since we have shown that the first resonance cannot be fully understood by finite-temperature theory, the applicability of the theory at large scattering lengths is unclear. In particular, the suppression of the ground state Efimov resonance is not understood, since it cannot be accounted for by universal zero-range theory. Another possible explanation is the presence of higher-body processes, which become significant compared to three-body recombination close to the resonance center. If four- or higher-body losses are more rapid than three-body losses, the three-body Efimov resonance is not visible. Moreover, the size $\sim \SI{0.7}{\micro m}$ of the excited state Efimov trimer may affect the dynamics in the trap, which on the smallest axis has a characteristic length of $\SI{1.7}{\micro m}$.

\section{Conclusion}
\label{sec:conclusion}

We have studied the ground-state Efimov resonance in \39K at various temperatures and our observations suggest that with decreasing temperature, the obtained value of $|a_-|$ becomes smaller, and the resonance becomes less prominent. The former is attributed to strong finite-range effects; the change in $a_-$ is far more dramatic than in similar measurements performed in Cs~\cite{Huang2015}. The flattening of the resonance is still an outstanding problem: the observed behavior arises due to effects not included in the simple zero-range model, e.g., finite-range effects or many-body physics, and theoretical calculations beyond the existing models are required to understand our data. Since the existing theoretical models do not describe our data, an ideal characterization of the temperature dependence is not possible, motivating further theoretical analysis.

Furthermore, we have performed measurements close to the Fesh\-bach resonance center to investigate the prospects of observing a second Efimov resonance. However, we do not observe any resonance feature connected to an excited Efimov state. We believe that this observation is connected to incomplete understanding of the observed first Efimov resonance: since the ground-state resonance is not fully understood, it is difficult to reliably make predictions about the excited states.

Our measurements show that certain aspects of few-body physics are yet to be understood, and encourage deeper investigations of finite-range and many-body effects on three-body loss measurements.

\begin{acknowledgments}
We thank Hans-Werner Hammer, Peder S{\"o}rensen, Chris Greene, Robin C\^{o}t\'e, Aksel Jensen, Nikolaj Zinner, and Felix Werner for useful discussions. We additionally thank Hans-Werner Hammer and Dmitry Petrov for providing us with the data for $s_{11}$ from Refs.~\cite{Platter2008, Rem2013} for benchmarking. L. J. W., N. B. J., K. T. S., M. G. S. and J. J. A. acknowledge support from the Villum Foundation, the Carlsberg Foundation, and the Danish Council for Independent Research. A. G. V. acknowledges support from the Humboldt Foundation.
\end{acknowledgments}

\bibliography{wacker_bib}

\end{document}